\documentclass[preprint,5p,twocolumn]{elsarticle}

\usepackage{graphicx,dcolumn,bm,amsmath,amssymb,mathrsfs,amsfonts,amsthm,tensor,braket,color,mathtools,autobreak,slashed,xfrac}
\usepackage{lineno}
\modulolinenumbers[5]
\usepackage{hyperref}
\hypersetup{colorlinks,bookmarksopen,bookmarksnumbered,citecolor=[rgb]{0.0,0.4,0.0},linkcolor=[rgb]{0.0,0.0,0.5},urlcolor=[rgb]{0.0,0.4,0.0},pdfstartview=FitH,linktocpage}

\def\dla{\langle\!\langle}
\def\dra{\rangle\!\rangle}
\def\be{\begin{enumerate}}
\def\ee{\end{enumerate}}
\def\bitem{\begin{itemize}}
\def\eitem{\end{itemize}}
\def\beq{\begin{equation}}
\def\eeq{\end{equation}}
\def\bea{\begin{eqnarray}}
\def\eea{\end{eqnarray}}

\def\D{\Delta}

\def\la{\langle}
\def\ra{\rangle}

\def\Mpl{M_{\textrm{P}}}

\def\3halfs{\textstyle{\frac{3}{2}}} 
\newcommand{\sla}[1]%
        {{\raise.15ex\hbox{$/$}\kern-.57em #1}}
\newcommand{\Sla}[1]%
        {{\raise.15ex\hbox{$/$}\kern-.75em #1}}

\journal{PLB, published as Phys. Lett. B 829 (2022) 137034, \url{https://doi.org/10.1016/j.physletb.2022.137034}}

\begin{document}
\hypersetup{colorlinks,bookmarksopen,bookmarksnumbered,citecolor=[rgb]{0.0,0.4,0.0},linkcolor=[rgb]{0.0,0.0,0.5},urlcolor=[rgb]{0.0,0.4,0.0},pdfstartview=FitH,linktocpage}

\begin{frontmatter}

\title{Testing Lorentz invariance of electrons with LHAASO observations of PeV gamma-rays from the Crab Nebula}

\author[pku]{Chengyi Li}
\author[pku,CHEP,CICQM]{Bo-Qiang Ma\corref{em}}
\cortext[em]{Corresponding author \ead{mabq@pku.edu.cn}}

\address[pku]{School of Physics and State Key Laboratory of Nuclear Physics and Technology, Peking University, Beijing 100871, China}
\address[CHEP]{Center for High Energy Physics, Peking University, Beijing 100871, China}
\address[CICQM]{Collaborative Innovation Center of Quantum Matter, Beijing, China}

\begin{abstract}
The Large High Altitude Air Shower Observatory~(LHAASO) recently reported the detection of gamma-ray emissions with energies up to $1.1~\textrm{PeV}$ from the Crab Nebula. Using the absence of vacuum Cherenkov effect by inverse-Compton electrons, we improve previous bounds to linear-order Lorentz invariance violation (LV) in the dispersion relations of electrons by $10^{4}$ times. We show that the LV effect on electrons is severely constrained, compatible with certain type of LV as expected by some models of quantum gravity~(QG), such as the string/D-brane inspired space-time foam. We argue that such models are supported by the Crab Nebula constraints from the LHAASO observations, as well as various LV phenomenologies for photons to date.
\end{abstract}

\begin{keyword}
PeV gamma-ray, Lorentz invariance violation, stringy space-time foam, Crab Nebula
\end{keyword}

\end{frontmatter}

Various attempts to new physics beyond the standard model and Einstein's relativity, especially those scenarios of quantum gravity~(QG), motivate the violations of Lorentz invariance~(LV) in the ultraviolet regime. Such violations may arise from various theories of QG such as stringy models for space-time foam~\cite{Ellis:1992eh}. The potentially measurable deviation from the Lorentz-invariant~(LI) physics is miniscule, but celestial objects of energetic emissions~\cite{Amelino-Camelia:1997ieq} provide nevertheless valuable probes of possible QG imprints on lower-energy world.

The Crab Nebula, as a bright source of $\gamma$-rays associated with a supernova explosion witnessed in 1054~A.D., is considered to be an ideal system to investigate LV. On 4 January 2021 at 16:45:06~UTC, the Kilometer Square Array~(KM2A) of the Large High Altitude Air Shower Observatory~(LHAASO) was triggered by photon flux from this object. Together with the Water Cherenkov Detector Array~(WCDA) measurement for a year duration, the collaboration recently reported the observation~\cite{Cao:2021pre1} of $\gamma$-rays with energies $\gtrsim 500$~GeV to $1.1$~PeV, implying the presence of an extreme particle accelerator in the nebula.

In this letter, we use this observation~\cite{Cao:2021pre1} to strengthen the bounds on~(superluminal) LV of $\mathcal{O}(E/\Mpl)$ for electrons, where $\Mpl=(\hbar c^5/G)^{1/2}\approx 1.22\times10^{19}~\textrm{GeV}$ is the Planck scale. By consideration of the vacuum Cherenkov effect by the $\gtrsim 1-2$~PeV electrons/positrons producing the highest-energy $\gamma$-ray event detected by LHAASO, we improve on the old Cherenkov constraint by four orders of magnitude. We show that the Lorentz violation for electrons is highly constrained~(in both superluminal and subluminal cases), indicating the absence of LV effects acting on electrons. This observation can serve as a support for the stringy space-time foam scenario, a string-theoretic framework compatible with various LV phenomenologies for photons to date~\cite{Li:2021gah}.

We describe possible LV deformations induced by QG in terms of modified particle dispersion relations, which can be parameterized, in a generic form, as
\beq
\label{eq:gdr}
E_{I}^{2}=m_{I}^{2}+p_{I}^{2}\Biggl[1+\delta_{I}^{(2)}-\sum_{n=1}^{\infty}s_{I,n}\biggl(\frac{p_{I}}{E_{\mathrm{LV}I,n}}\biggr)^{n}\Biggr],
\eeq
where $(E_{I},\mathbf{p}_{I})=P_{I}$ is the 4-momentum, $p_{I}=\lvert\mathbf{p}_{I}\rvert$, and $E_{\textrm{LV}I,n}$ is the $n$th-order LV scale for particle $I$~($=\gamma,e_{\textrm{R,L}}^{\mp}$, with the label ``R'' or ``L'' indicating positive or negative helicity mode). The sign factor, $s_{I,n}$, refers to superluminal~($s _{I,n}=-1$), and subluminal~($s_{I,n}=+1$) dominant phenomena. Units with $\hbar=1$ and the low energy speed of light $c=1$ are adopted. We allow a $p_{I}^{2}\delta_{I}^{(2)}$ term in~(\ref{eq:gdr}) of~$I=e$. The bound $\lvert\delta_{e}^{(2)}\rvert<2\times 10^{-15}$ is provided by analyses of synchrotron energy losses at LEP~\cite{Altschul:2010na}.

For $m_{I}\ll E_{I}\ll\Mpl$, the $p_{I}^{n+2}~(n>0)$ LV terms are supposed to dominate. Assume that $v_{I}=\partial E_{I}/\partial p_{I}$ still holds in QG, the propagation velocity of a particle, to first order in $m_{I}^{2}$, can be computed as
\beq
\label{eq:gps}
v_{I}\simeq 1-\sfrac{1}{2}\biggl(\frac{m_{I}}{p_{I}}\biggr)^{2}-s_{I,n}\frac{1+n}{2}\biggl(\frac{p_{I}}{E_{\textrm{LV}I,n}}\biggr)^{n},
\eeq
where the linear~($n=1$) or quadratic~($n=2$) LV term is usually considered. We eliminate at this stage the $\mathcal{O}(p_{I}^{2})$ term in~(\ref{eq:gdr}), otherwise an extra term $(\sfrac{1}{2})\delta_{e}^{(2)}$ would enter in the group velocity~(\ref{eq:gps}) of electrons/positrons.\footnote{We shall come back to that later, but for now $\delta_{e}^{(2)}$ is forced to be null since it has been tightly constrained, as mentioned.} We assume that the conservation laws of energy and momentum remain intact, as we adopt here a purely phenomenological formalism of LV~(cf.,~(\ref{eq:gdr})), lacking a detailed knowledge from the underlying QG.

With modified dispersion relations~(\ref{eq:gdr}), peculiar new phenomena beyond the standard model~(SM) emerge. For example, Cherenkov radiation of leptons, $e\rightarrow e\gamma$, may occur in a vacuum, and synchrotron emission in magnetic fields may also be modified. For photons, the universality of their velocities would no longer hold, and the spontaneous self-decay of a photon into a $e^{+}e^{-}$ pair may happen. Astrophysical objects of high energy emissions~\cite{Amelino-Camelia:1997ieq}, such as the Crab Nebula, have proven ideal laboratories for exploring these LV effects.

The spectrum of the Crab Pulsar Wind Nebula~(PWN) consists of two marked humps as described by the~(one-zone) synchro-self-Compton model~\cite{Atoyan:1996hgu,Meyer:2010tta}, according to which the emission from radio to low frequency gamma-rays comes from synchrotron radiation by relativistic electrons/positrons, while high energy $\gamma$-rays above $\sim 1$~GeV are produced via inverse-Compton~(IC) process by these electrons. Precise measurement of the Crab spectra allows one to testify the standard LI prediction of the radiation properties and, eventually, to probe LV. The most severe of all known limits to electron Lorentz violation~(i.e., $E_{{\textrm{LV}e,n}}$) does come from the Crab Nebula~(see below for more details), while in the following we further constrain it via IC Cherenkov effect by using the latest LHAASO observation.

As mentioned, the dispersion relation~(\ref{eq:gdr}) implies that a free electron in the vacuum can emit a Cherenkov photon~\cite{Jacobson:2005bg}. The emitted photon is soft enough so that its LV can be neglected. This process occurs once the electron group velocity reaches the low energy speed of light, $v_{e}\approx 1$, then the threshold can be computed as
\beq
\label{eq:VCc}
\frac{m_{e}^{2}}{(1+n)\lvert\mathbf{p}_{e}\rvert^{n+2}}\leq\frac{-s_{e,n}}{E_{\mathrm{LV}e,n}^{n}},
\eeq
for any order $n$, with $m_{e}$ the electron rest mass.

We focus on LV suppressed by one power of the ratio $E/\Mpl$. Therefore, the threshold~(\ref{eq:VCc}) reduces to
\beq
\label{eq:VCth}
E_{e}\gtrsim\biggl(-\frac{m_{e}^{2}E_{\textrm{LV}e}}{2s_{e}}\biggr)^{1/3},
\eeq
with $p_{e}\simeq E_{e}$ for high energies~(i.e., $E_{e}\gg m_{e}$).\footnote{We omit the subscript $n$ in the case of $n=1$.} The threshold~(\ref{eq:VCth}) is meaningful only if $s_{e}=-1$, such that any energetic electrons above $E_{e,\textrm{th}}\simeq(m_{e}^{2}E_{\textrm{LV}e}/2)^{1/3}$ would dissipate their energy extremely quickly~\cite{Jacobson:2002td}. The rate of energy loss is orders of magnitude greater than that due to the IC scattering. Therefore, relativistic electrons must be stable in order to produce the ultrahigh-energy~(UHE) $\gamma$-rays inside the PWN. Observations of these $\gamma$-rays can hence be used to constrain the scale of possible \textit{superluminal} Lorentz violation, $E_{\textrm{LV}e}^{(\textrm{sup})}$, for electrons.

A previous Cherenkov bound, $E_{\textrm{LV}e}^{(\textrm{sup})}\gtrsim 10^{21}$~GeV, was obtained~\cite{Jacobson:2003bn} using the existence of 50~TeV electrons inferred via the detection of 50~TeV photons~\cite{Tanimori:1998ks} from the Crab Nebula. Electrons of energy up to 75~TeV are necessary to explain the later HEGRA's observations of 75 TeV $\gamma$-rays~\cite{HEGRA:2004tpc}. This then strengthens the constraint by a factor of $(3/2)^{3}\approx 3$. Due to the collaborations Tibet-AS$\gamma$~\cite{Amenomori:2019rjd}, HAWC~\cite{HAWC:2019xdx} and LHAASO~\cite{Aharonian:2020iou}, the IC $\gamma$-spectra of the Crab that continue beyond 100~TeV have been measured since 2019. In particular, photons observed by the Tibet-AS$\gamma$~\cite{Amenomori:2019rjd} up to $\sim 450$~TeV indicate that the parent electrons may reach sub-PeV energies, we deduce for this case that $E_{e}\geq 450$~TeV as a conservative estimate, or that $E_{\textrm{LV}e}^{(\textrm{sup})}>2E_{e}^{3}/m_{e}^{2}\sim 7\times 10^{23}$~GeV.

Recently, LHAASO-KM2A reported the first detection of $\gamma$-rays of energy exceeding 1~PeV from the Crab Nebula~\cite{Cao:2021pre1}. The most energetic event of $1.12\pm 0.09$~PeV was registered. This implies that the IC spectrum of the Crab extends at least out to $\sim 1.1$~PeV. Assuming an electronic origin of these $\gamma$-rays~\cite{Cao:2021pre1}, a stronger constraint can then arise from the stability of $\gtrsim 1$~PeV electrons against Cherenkov effect.

We now derive this Cherenkov bound by taking into account more precisely the energies of parent electrons. Given that the IC scattering is basically unaffected by LV~(see Appendix), its kinematics can be described with standard means of special relativity. In this context, the energies $E_{\gamma}$ of produced $\gamma$-rays via Compton interactions of the ambient target photons~(of energy $\omega_{t}$) with relativistic electrons~(or positrons) lie in the region:
\beq
\label{eq:Epr}
\left(1+\frac{1}{b_{\textrm{KN}}}\right)^{-1}E_{e}\geq E_{\gamma}\gg\omega_{t},
\eeq
where $b_{\textrm{KN}}=4\omega_{t}E_{e}/m_{e}^{2}$ is the Klein-Nishina~(KN) parameter for the IC process. This yields
\beq
\label{eq:Ees}
E_{e}\equiv\frac{E_{\gamma}}{\mathcal{Z}}\geq\frac{1}{2}E_{\gamma}+\frac{\sqrt{\D(E_{\gamma};\omega_{t})}}{2\omega_{t}},
\eeq
where $\mathcal{Z}=E_{\gamma}/E_{e}\leq b_{\textrm{KN}}/(1+b_{\textrm{KN}})$ denotes the energy fraction in case of single scattering, and $\D(E_{\gamma};\omega_{t})=\omega_{t}E_{\gamma}(m_{e}^{2}+\omega_{t}E_{\gamma})$. In the Crab Nebula, several radiation fields supply target photons for IC reactions. However, at energies $\gtrsim 100$~TeV, the 2.7~K cosmic microwave background~(CMB) field dominates the $\gamma$-ray production~\cite{Atoyan:1996hgu,Meyer:2010tta}. For the LHAASO 1.1~PeV event, on noting that $\omega_{t(=\textrm{CMB})}\sim 6\times 10^{-4}~\textrm{eV}$, $\sfrac{m_{e}^{2}}{\omega_{t}E_{\gamma}}\sim\mathcal{O}(.1)$, the lower energy limit, $E_{e,0}\leq E_{e}$, for parent electrons is
\beq
\label{eq:Eee}
E_{e,0}=E_{\gamma}+\frac{m_{e}^{2}}{4\omega_{\textrm{CMB}}}+\mathcal{O}\biggl(\frac{m_{e}^{4}}{\omega_{\textrm{CMB}}^{2}E_{\gamma}}\biggr),
\eeq
where the leading contribution is factored out rather than the general kinematic condition~(\ref{eq:Epr}) for computational convenience. Eq.~(\ref{eq:Eee}) yields $E_{e}\gtrsim 1.2~\textrm{PeV}$.~(A stronger IC Cherenkov constraint could clearly be obtained by inserting this $E_{e,0}$ in $E_{e}\leq E_{e,\textrm{th}}$.)

\begin{figure}[t]
\centering
\includegraphics[scale=0.32]{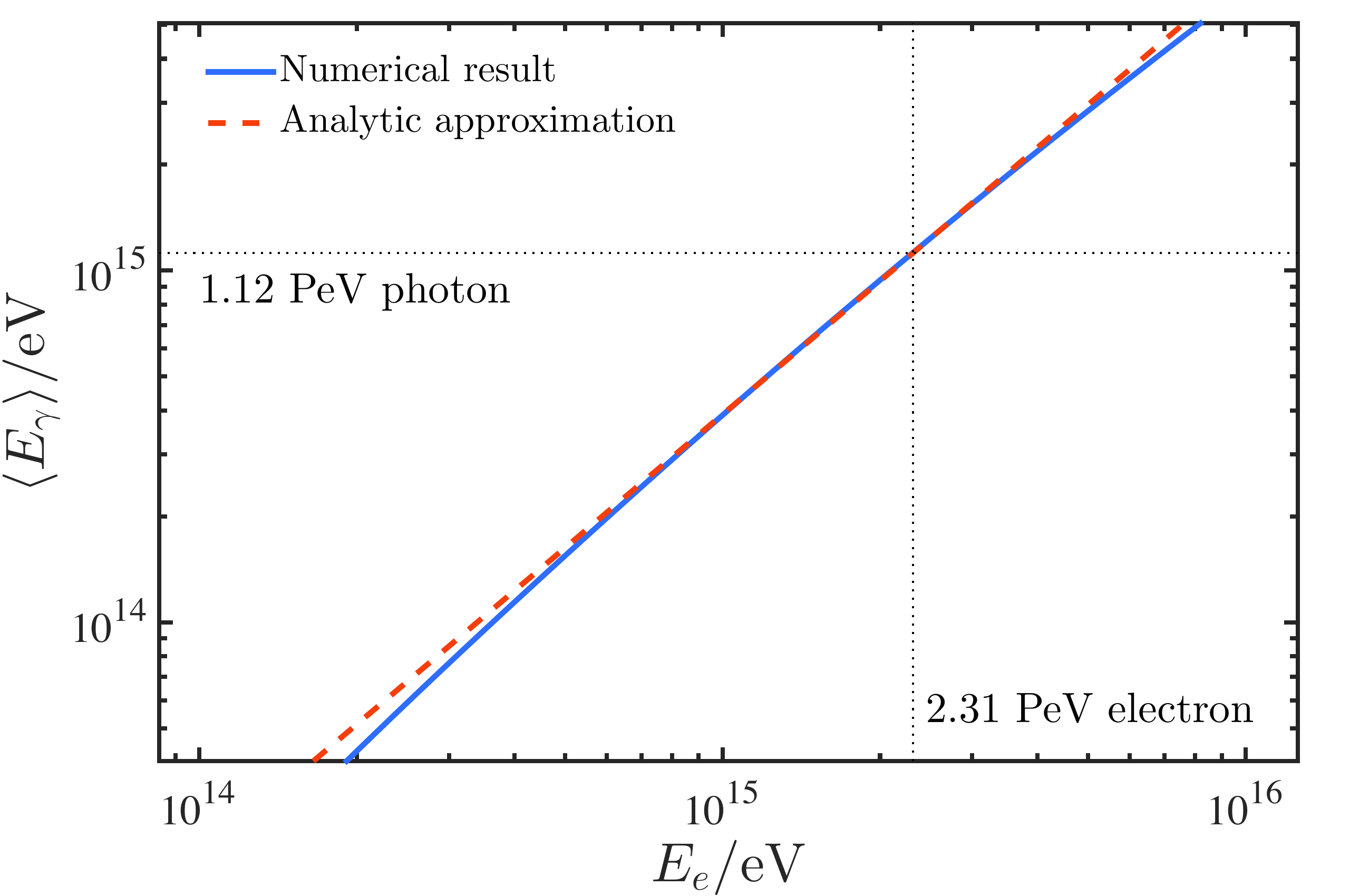}
\caption{Curves of energies of IC $\gamma$-rays with respect to that of parent electrons. The solid curve denotes the numerical result of Eq.~(\ref{eq:Epoz}), whilst the dashed coloured curve shows the analytic expression that we adopt, i.e., $\langle E_{\gamma}\rangle\simeq 0.39(E_{e}/\textrm{PeV})^{1.26}$~PeV, which within the interval $0.3~\textrm{PeV}\leq E_{\gamma}\leq 3$~PeV provides an accuracy $\gtrsim 2\%$, higher than that achieved in Ref.~\cite{Cao:2021pre1}.}
\label{fig:er}
\end{figure}

In fact we can do better by integration over the Planckian distribution of the~(isotropic) CMB field, rather than using the single scattering result as presented above. In such a case, the energy spectrum of the up-scattered photons can be estimated via an approximate analytical calculation~\cite{Khangulyan:2013hwa}. As $\gamma$-ray  production above 100~TeV proceeds in the KN regime~($b_{\textrm{KN}}\gg 1$), the average energy $\la E_{\gamma}\ra$ of the IC $\gamma$-rays can be evaluated  as~\cite{Khangulyan:2013hwa}
\beq
\label{eq:Epoz}
\langle E_{\gamma}\rangle=\langle\mathcal{Z}\rangle_{\textrm{CMB}}E_{e}=\frac{4E_{e}\mathcal{T}}{4\mathcal{T}+0.3}\frac{\ln(1+\mathcal{T})}{\ln(1+4\mathcal{T})},
\eeq
where $\mathcal{T}=k_{\textrm{B}}T_{\textrm{CMB}}E_{e}/m_{e}^{2}$, with $k_{\textrm{B}}$ the Boltzmann constant, $T_{\textrm{CMB}}\simeq 2.7$~K. By adopting an analytic approximation of~(\ref{eq:Epoz}), $E_{\gamma}\simeq 0.39(E_{e}/\textrm{PeV})^{1.26}$~PeV~(see Fig.~\ref{fig:er}), we infer that a $\sim 2.3$~PeV electron is required to produce 1.1~PeV photon~\cite{Cao:2021pre1} as reported by LHAASO.

The absence of the Cherenkov threshold~(\ref{eq:VCth}) up to such energies then yields a constraint of
\beq
\label{eq:VCb}
E_{\textrm{LV}e}^{(\textrm{sup})}\gtrsim 7.66\times 10^{33}~\textrm{eV}\left(\frac{E_{e}}{\textrm{PeV}}\right)^{3},
\eeq
or equivalently $E_{\textrm{LV}e}\gtrsim 7.21\times 10^{25}(E_{\gamma}/\textrm{PeV})^{2.38}$~GeV for $s_{e}=-1$ as considered. For the highest-energy event from the Crab Nebula~\cite{Cao:2021pre1} of $E_{\gamma,\max}=1.12$~PeV therefore,
\beq
\label{eq:VCb_n}
E_{\textrm{LV}e}^{(\textrm{sup})}\gtrsim 9.4\times 10^{25}~\textrm{GeV},
\eeq
over about 7,500,000 times the Planck scale. Taking into account the uncertainty $\lvert\delta E_{\gamma,\max}\rvert=0.09$~PeV the above result would be $9.44_{-1.71}^{+1.91}\times 10^{25}$~GeV. This improves the previous IC Cherenkov limit~\cite{Jacobson:2002td,Jacobson:2003bn} by a factor $\gtrsim 10^{4}$. We infer in passing a constraint $E_{\textrm{LV}e,2}^{(\textrm{sup})}>\sqrt{3}E_{e}^{2}/m_{e}\sim 2\times 10^{16}$~GeV for $n=2$ LV effects in Eq.~(\ref{eq:gdr}).

Our result~(\ref{eq:VCb_n}) can also be used to impose stronger limits on those theories entailing LV for electrons/positrons. Applying it, e.g., to the dimension-5 standard-model extension~(SME)~\cite{Myers:2003fd}, the relevant LV parameter is constrained~(from the positive side) by $\eta_{(\textrm{R/L})}\lesssim1.3\times 10^{-7}$. This largely improves the previous bounds~($\mathcal{O}(10^{-2})$)~\cite{Jacobson:2003bn} on a positive $\eta$ from the IC Cherenkov effect.

We obtain~(\ref{eq:VCb_n}) under the assumption $\delta_{e}^{(2)}=0$, as mentioned. Allowing for its non-zero values does not significantly affect our result. Further, we can estimate a bound to that parameter by comparing its contribution to group velocities with that due to a linear-order Planck suppressed term, i.e., $(\sfrac{1}{2})\delta_{e}^{(2)}\approx v_{e}-1\simeq E_{e}/E_{\textrm{LV}e}$. This yields
\beq
\label{eq:deltave}
\delta_{e}^{(2)}\lesssim 2\Bigl(E_{e,\max}/E_{\textrm{LV}e}^{(\textrm{sup})}\Bigr)\simeq 4.9\times 10^{-20},
\eeq
given that $E_{e,\max}\simeq 2.31$~PeV. Though this result is only an estimate but agrees surprisingly with the rigorous limit, $\delta_{e}^{(2)}\leq (E_{e,\max}/m_{e})^{-2}\simeq 5\times 10^{-20}$, as derived by using a careful analysis of, e.g., Ref.~\cite{Altschul:2006pv}. This new constraint~(\ref{eq:deltave}) is more than five orders of magnitude stronger than the best laboratory constraint~\cite{Altschul:2010na}, and is comparable to that constraint from the 2010 Crab flare~\cite{Stecker:2013jfa}~(with an improvement by a factor of $\simeq 5$). [N.B.: $\delta_{e}^{(2)}$ adopted here is twice that $\delta_{e}$ defined in~\cite{Stecker:2013jfa}.] This result~(\ref{eq:deltave}) applies directly to, e.g., the Coleman-Glashow model~\cite{Coleman:1998ti}.

As an aside, we note that the latest detection of PeV photons from the Crab Nebula at LHAASO~\cite{Cao:2021pre1} can also be used to limit photon decay, $\gamma\rightarrow e^{+}e^{-}$. Such process, as made possible by superluminal Lorentz violation in photon dispersions~(\ref{eq:gdr}), would also result in the non-observation of these Crab $\gamma$-rays at Earth. A stringent bound on Lorentz violation for photons~(in the superluminal case) is therefore provided by the measured $\gamma$-rays with energies up to the highest at $E_{\gamma,\max}$. Applying the decay threshold equation~\cite{Li:2021tcw}, we have
\begin{align}
\label{eq:pdb_n1}
E_{\textrm{LV}\gamma}>&\ E_{\gamma,\max}^{3}/4m_{e}^{2}\simeq 1.35^{+0.35}_{-0.30}\times 10^{24}~\textrm{GeV},\nonumber\\
&\qquad\textrm{for}~n=1,~s_{\gamma}=-1.
\end{align}
This result can be directly transferred to the limits on certain models of LV, leading to, e.g., $\xi\lesssim 10^{-6}\ll\mathcal{O}(1)$ for the SME with dimension-5~($\mathcal{CPT}$-odd) operators~\cite{Myers:2003fd}. Incidentally, we deduce for the quadratic-order LV terms that $E_{\textrm{LV}\gamma,2}^{(\textrm{sup})}>E_{\gamma,\max}^{2}/2m_{e}\simeq 1.23^{+0.21}_{-0.19}\times 10^{15}~\textrm{GeV}$. These $\gamma$-decay constraints are comparable to the existing bounds as obtained recently in Refs.~\cite{Li:2021tcw,Chen:2021hen,Cao:2021pre2} based on the previous discovery of 12 Galactic PeVatrons~\cite{Aharonian:2021pre} by the LHAASO collaboration.

We emphasize again that our constraints, as derived above, are only applicable to the superluminal LV scenarios~(i.e., $s_{e/\gamma}=-1$). For the \textit{subluminal} case, a complementary constraint for electrons, $E_{\textrm{LV}e}^{(\textrm{sub})}\gtrsim 10^{24}$~GeV, was deduced~\cite{Maccione:2007yc} at $95\%$ confidence, by utilizing the highest-frequency~($\sim 0.1$~GeV) synchrotron radiation in the Crab Nebula. Remarkably, the new Cherenkov constraint~(\ref{eq:VCb_n}) reaches the order of magnitude~(with a mild excess of $\mathcal{O}(100)$) the tightest synchrotron constraint.

All these constraints severely limit the linear-order LV corrections for electrons/positrons, making any theory that predicts such type of LV very unlikely. Meanwhile, the superluminal dominate effect on photons has also been tightly bounded. These facts may enable one to exclude any possibility of linear deformations of LV. However, we indicate here that this is not the case. The analysis leading to above constraints, at least, has nothing to do with the magnitude of the \textit{subluminal} Lorentz violation for \textit{photons}, and as such there is still space for this type of new physics to live. For example, the Lorentz-violating picture for photons, as suggested from gamma-ray burst~(GRB)~\cite{Shao:2009bv,Zhang:2014wpb,Xu:2016zan,Liu:2018qrg,Zhu:2021pml} or active galactic nucleus~(AGN) data~\cite{Li:2020uef}, may still be viable~\cite{Amelino-Camelia:2016ohi}.

In effect, on the other hand, certain models of QG originating from string/brane theory could account for all of these phenomenologies. Such models have been termed stringy space-time foam~\cite{Ellis:2004ab,Ellis:2008gg,Li:2009tt,Li:2021eza}, where solitonic defects, such as pointlike D0-branes~(``D-particles'') allowed in type I/IIA strings, populate the higher-dimensional bulk space, and give the background space-time a foamy nature~(``space-time foam'' or \textit{D-foam}). The presence of such defects breaks locally Lorentz symmetry. Standard-model particles represented as~(open-)string excitations confined to the~(compactified) three-brane~(describing the observable Universe) may interact topologically nontrivially with these stringy defects.

The interactions of photon open-string states with D-foam defects, necessitating the emission of nonlocal intermediate strings with the momentum transfer $\Delta\mathbf{p}$, lead to a LV vacuum that could slow down the high energy photons in terms of a target-space metric disturbance, $g_{\mu\nu}=\eta_{\mu\nu}+h_{\mu\nu}$. Here $h_{\mu\nu}$ is related to the D-defect recoil 3-velocity $\mathbf{u}$ via $h_{0i}=u_{i}(1-\mathbf{u}^{2})^{-1/2}\simeq (g_{s}/M_{s})\Delta p_{i}\equiv\lambda p_{\gamma i}\equiv g_{s}\zeta p_{\gamma i}/M_{s}$~\cite{Ellis:2004ab,Li:2021eza}. Averaging it over both statistical effects and quantum effects~(denoting such a double average by $\dla\cdots\dra$), and assuming an anisotropic D-foam situation, $\dla\lambda\dra\neq 0$, this yields a modified relativistic relation of radiation on a D-brane world as
\beq
\label{eq:sdr}
\mathbf{p}_{\gamma}^{2}-E_{\gamma}^{2}\simeq\varsigma_{\textrm{D}}\ell_{s}\mathbf{p}_{\gamma}^{2}E_{\gamma},
\eeq
where the correction term scales \textit{linearly} with the energy. $\ell_{s}$ is the string scale and $\varsigma_{\textrm{D}}$ is a QG parameter~\cite{Li:2021eza} related to the mass scale of D-particles, $M_{\textrm{D}}= M_{s}/g_{s}$, via
\beq
\label{eq:qgp}
\varsigma_{\textrm{D}}=\frac{\dla2\zeta\dra}{M_{\textrm{D}}\ell_{s}}\equiv 2g_{s}\zeta_{\textrm{D}},~~M_{s}=\ell_{s}^{-1},
\eeq
with $\zeta_{\textrm{D}}<1$ an $\mathcal{O}(1)$ parameter and $g_{s}<1$ is the~(weak) string coupling. From Eq.~(\ref{eq:sdr}), one would expect a \textit{subluminal} photon propagation velocity \textit{in vacuo}:
\beq
\label{eq:pgv}
v_{\gamma}^{\textrm{D-foam}}=1-\biggl(\frac{\varsigma_{\textrm{D}}}{M_{s}}\biggr)E_{\gamma}+\ldots,~~\varsigma_{\textrm{D}}>0.
\eeq

In contrast to photons, there is however \textit{no} analogous interaction between charged fermionic strings with D-particles, due to the electric charge flux conservation. This is traced back to the facts that~\cite{Ellis:2003ia,Ellis:2003sd} excitations that are charged under the gauge group~(i.e., $U(8)\times U(8)^{\prime}$ in models of~\cite{Ellis:2004ab,Ellis:2008gg,Li:2021gah} as considered here) fall into bifundamental representations, and that D-particles are~(electrically) neutral, as are its excitations, which transform in the adjoint representation, hence behave like gauge particles. For an incident open-string state that carries electric flux, the formation of nonlocal intermediate string states during collisions with D-defects is thus not allowed. This fact leads to a transparency of the D-particle space-time foam to charged open-string excitations. Electrons coupling to a space-time background without exotic local curvatures related to $h_{\mu\nu}$ possess at tree level standard relativistic relations, $\dla P_{e}^{\mu}g_{\mu\nu}P_{e}^{\nu}\dra\simeq -E_{e}^{2}+\mathbf{p}_{e}^{2}=-m_{e}^{2}$, which are protected by special gauge symmetries in string theory, as explained, and as such, electrons would propagate in a Lorentz-invariant way. They \textit{do not} emit any Cherenkov radiation in vacuum.

Note that our strong constraint~(\ref{eq:VCb_n}) just indicates that the vacuum Cherenkov process does not happen for the UHE electrons which produce the PeV IC radiation as observed by LHAASO. This coincides with the above model prediction. Since the strength of the constraints is improved by many orders of magnitude beyond the Planck scale, such~(stringy) models in which no LV is present for electrons/positrons or in general for charged leptons are therefore supported by these gradually tightened constraints from the Crab Nebula.

The severe bound, e.g.,~(\ref{eq:pdb_n1}), from $\gamma$-decays is also consistent with this type of D-brane foam models. Since the existence of D-particles on the branes leads to a deceleration of LHAASO $\gamma$-rays, corresponding to a subluminal propagation of light. Photons are therefore \textit{stable}, implying that the self decay \textit{will not} happen for arbitrarily attainable high energies. This agrees with those stringent limits, by noting that PeV photons do survive without decaying away. The string/brane foam models thus gain additional supports from those $\gamma$-decay limits.

It is worthy noticing that many other astrophysical results/constraints on LV, available today, such as the subluminal light speed variation~\cite{Shao:2009bv,Zhang:2014wpb,Xu:2016zan,Liu:2018qrg,Zhu:2021pml,Li:2020uef,Amelino-Camelia:2016ohi} suggested previously from GRB/AGN data, can be simultaneously explained with such QG scenarios. For a more extensive and up-to-date discussion on this argument, we refer the reader to Refs.~\cite{Li:2021gah,Li:2021tcw}, see also~\cite{Li:2021eza,Li:2021duv}.

We should stress at this stage that it is still premature to think of any violations of Lorentz invariance from QG as essential ingredients of nature. If, however, future observational $\gamma$-ray data for the Crab Nebula break further the existing record of the highest photon energy, the space-time foam models associated with string theory/M-theory and supersymmetry would get stronger supports as one of the most promising models of QG that can fit the astrophysical observations.

In this letter, we place constraints on Lorentz violation by exploiting the newly reported LHAASO 1.1~PeV event, which turns out to be the most high-energetic photon coming from the ancient Crab Nebula. We obtain for superluminal electron dispersions, $E_{\textrm{LV}e}\gtrsim 10^{26}$~GeV, which further excludes the linear-order LV effect on electrons~(or positrons). This constraint improves the old IC Cherenkov bound~\cite{Jacobson:2002td,Jacobson:2003bn} by at least $10^{4}$ times. Based upon various phenomenological results on LV to date, we indicate that certain models of QG, especially some string theory frameworks only introducing LV for charge-neutral excitations such as photons but not for particles like electrons, might gain supports from the new constraints imposed by the Crab Nebula observation.

\section*{Acknowledgments}

This work is supported by National Natural Science Foundation of China (Grant No.~12075003).

\section*{Appendix A. Comments on IC processes}
\label{app}
\numberwithin{equation}{section}
\renewcommand{\theequation}{A.\arabic{equation}}
\setcounter{equation}{0}

In the Lorentz-invariant case, the energies of inverse-Compton~(IC) photons can be worked out, by using the relativistic parallax formula and the usual kinematics of Compton scattering, as~(in units of $m_{e}=1$),
\beq
\label{eq:a1}
E_{\gamma}^{\mathcal{O}}=\widetilde{\gamma}^{2}\omega_{t}^{\mathcal{O}}\frac{(1-\cos\phi^{\mathcal{O}})(1+\cos\theta^{\mathcal{E}}\!\cos\phi^{\mathcal{E}})}{1+\widetilde{\gamma}\omega_{t}^{\mathcal{O}}(1-\cos\phi^{\mathcal{O}})(1-\cos\theta^{\mathcal{O}})}.
\eeq
The Lorentz factor $\widetilde{\gamma}$ is that of electrons, $\theta$ is the scattering angle and $\phi$ is the angle between the initial photon and electron. Superscripts denote quantities taken in the rest frame of the observer~($\mathcal{O}$) or of the electron~($\mathcal{E}$). Consider head-on collisions where ultrarelativistic electrons transfer substantial fractions of their energy to photons, $v_{e}\sim 1$, $\theta,\phi\approx\pi$, so $E_{\gamma}=b_{\textrm{KN}}E_{e}/(1+b_{\textrm{KN}})$.

Note that these calculations involve the boost of the kinematics between frames, whereas the transformation law is unclear in the presence of Lorentz violation. We now examine the question of neglecting LV terms in evaluating the IC photon energy. Let us work fully in the observer's frame, and assume that the photon dispersion is subluminal while the electron LV scale $\ell_{e}(\equiv s_{e}/E_{\textrm{LV}e})$ is negative. The conservation law reads
\beq
\label{eq:a2}
P_{e}+P_{t}-P_{\gamma}=P_{e}^{\prime},
\eeq
where $P_{e}^{\prime}$ is the 4-momentum for the electron after interaction. Let $Q$ be the 4-vector recoil momentum $Q=P_{t}-P_{\gamma}$. Taking the square of Eq.~(\ref{eq:a2}) we get
\beq
\label{eq:a3}
\widetilde{v}\mathcal{Q}\cos\chi\approx 1-x-\biggl[1-\biggl(1+\frac{1}{2}x\ell_{\gamma}\omega_{t}\biggr)\!\cos\theta\biggr]\mathcal{Z}\equiv F(x;\theta),
\eeq
where higher-order terms are dropped given that $\ell_{e/\gamma}\sim\mathcal{O}(\Mpl^{-1})$. $\chi$ is an angle between 3-vector $\mathbf{q}=\mathbf{p}_{t}-\mathbf{p}_{\gamma}$ and $\mathbf{p}_{e}$, $\ell_{\gamma}\equiv s_{\gamma}/E_{\textrm{LV}\gamma}$, $x\equiv E_{\gamma}/\omega_{t}$, $\widetilde{v}\equiv\lvert\mathbf{p}_{e}\rvert/E_{e}$,
\beq
\label{eq:a4}
\mathcal{Q}\equiv\frac{\lvert\mathbf{q}\rvert}{\omega_{t}}\simeq\sqrt{1+x^{2}(1+x\ell_{\gamma}\omega_{t})-2x\biggl(1+\frac{1}{2}x\ell_{\gamma}\omega_{t}\biggr)\!\cos\theta}.
\eeq
The limits of $x$~(or of $E_{\gamma}$) are defined from Eq.~(\ref{eq:a3}) at extreme values of $\cos\chi(=\pm 1)$ as
\beq
\label{eq:a5}
\frac{1}{E_{e}^{2}}\leq 1-\biggl(\frac{F}{\mathcal{Q}}\biggr)^{2}(x;\pi)+\mathcal{O}(\ell_{e}E_{e}).
\eeq
Solving this inequality one would find the maximal value of $E_{\gamma}$ in the Lorentz violation context. The exact solution may be tedious and complicated, whereas, it is unimportant. What matters is that at the kinematic level all LV terms intervene at a level of $<10^{-13}$ at $E\lesssim 1$~PeV, and one finds that the two most important features of IC scattering still hold: the $\gamma$-radiation is almost beamed alone the direction of the initial electrons~($\theta\approx\pi$), and the most-energetic IC photons can carry off energies almost at that of electrons~($E_{\gamma}\sim E_{e}$). Hence the IC reaction is hardly affected by LV, and neglecting terms of the order of $E\ell_{e/\gamma}$ with respect to unity is justified.



\begin{thebibliography}{99}

\bibitem{Ellis:1992eh}
J.~Ellis, N.~E.~Mavromatos, D.~V.~Nanopoulos,
{Phys. Lett. B} 293 (1992) 37.

\bibitem{Amelino-Camelia:1997ieq}
G.~Amelino-Camelia, J.~Ellis, N.~E.~Mavromatos, D.~V.~Nanopoulos, S.~Sarkar,
{Nature} 393 (1998) 763.

\bibitem{Cao:2021pre1}
Z.~Cao, F.~A.~Aharonian, Q.~An et al. [LHAASO Collaboration],
{Science} 373 (2021) 425.

\bibitem{Li:2021gah}
C.~Li, B.-Q.~Ma,
{Phys. Lett. B} 819 (2021) 136443.

\bibitem{Altschul:2010na}
B.~Altschul,
{Phys. Rev. D} 82 (2010) 016002.

\bibitem{Atoyan:1996hgu}
A.~M.~Atoyan, F.~A.~Aharonian,
{Mon. Not. R. Astron. Soc.} 278 (1996) 525.

\bibitem{Meyer:2010tta}
M.~Meyer, D.~Horns, H.-S.~Zechlin,
{Astron. Astrophys.} 523 (2010) A2.

\bibitem{Jacobson:2005bg}
T.~Jacobson, S.~Liberati, D.~Mattingly,
{Annals Phys.} 321 (2006) 150.

\bibitem{Jacobson:2002td}
T.~Jacobson, S.~Liberati, D.~Mattingly,
{Phys. Rev. D} 66 (2002) 081302(R);
ibid. 67 (2003) 124011.

\bibitem{Jacobson:2003bn}
T.~Jacobson, S.~Liberati, D.~Mattingly, F.~W.~Stecker,
{Phys. Rev. Lett.} 93 (2004) 021101.

\bibitem{Tanimori:1998ks}
T.~Tanimori et al.,
{Astrophys. J.} 492 (1998) L33.

\bibitem{HEGRA:2004tpc}
F.~A.~Aharonian et al. [HEGRA Collaboration],
{Astrophys. J.} 614 (2004) 897.

\bibitem{Amenomori:2019rjd}
M.~Amenomori et al. [Tibet AS$\gamma$ Collaboration],
{Phys. Rev. Lett.} 123 (2019) 051101.

\bibitem{HAWC:2019xdx}
A.~U.~Abeysekara et al. [HAWC Collaboration],
{Astrophys. J.} 881 (2019) 134;
{Phys. Rev. Lett.} 124 (2020) 021102.

\bibitem{Aharonian:2020iou}
F.~A.~Aharonian et al. [LHAASO Collaboration],
{Chin. Phys. C} 45 (2021) 025002.

\bibitem{Khangulyan:2013hwa}
D.~Khangulyan, F.~A.~Aharonian, S.~R.~Kelner,
{Astrophys. J.} 783 (2014) 100.

\bibitem{Myers:2003fd}
R.~C.~Myers, M.~Pospelov,
{Phys. Rev. Lett.} 90 (2003) 211601.

\bibitem{Altschul:2006pv}
B.~Altschul,
{Phys. Rev. D} 74 (2006) 083003.

\bibitem{Stecker:2013jfa}
F.~W.~Stecker,
{Astropart. Phys.} 56 (2014) 16.

\bibitem{Coleman:1998ti}
S.~R.~Coleman, S.~L.~Glashow,
{Phys. Rev. D} 59 (1999) 116008.

\bibitem{Li:2021tcw}
C.~Li, B.-Q.~Ma,
{Phys. Rev. D} 104 (2021) 063012.

\bibitem{Chen:2021hen}
L.~Chen, Z.~Xiong, C.~Li, S.~Chen, H.~He,
{Chin. Phys. C} 45 (2021) 105105.

\bibitem{Cao:2021pre2}
Z.~Cao, F.~A.~Aharonian, Q.~An et al. [LHAASO Collaboration], 
{Phys. Rev. Lett.} 128 (2022) 051102.

\bibitem{Aharonian:2021pre}
Z.~Cao, F.~A.~Aharonian, Q.~An et al. [LHAASO Collaboration],
{Nature} 594 (2021) 33.

\bibitem{Maccione:2007yc}
L.~Maccione, S.~Liberati, A.~Celotti, J.~G.~Kirk,
{J. Cosmol. Astropart. Phys.} {10} (2007) 013.

\bibitem{Shao:2009bv}
L.~Shao, Z.~Xiao, B.-Q.~Ma,
{Astropart. Phys.} 33 (2010) 312.

\bibitem{Zhang:2014wpb}
S.~Zhang, B.-Q.~Ma,
{Astropart. Phys.} 61 (2015) 108.

\bibitem{Xu:2016zan}
H.~Xu, B.-Q.~Ma,
{Astropart. Phys.} 82 (2016) 72;
{Phys. Lett. B} 760 (2016) 602;
{J. Cosmol. Astropart. Phys.} {01} (2018) 050.

\bibitem{Liu:2018qrg}
Y.~Liu, B.-Q.~Ma,
{Eur. Phys. J. C} 78 (2018) 825.

\bibitem{Zhu:2021pml}
J.~Zhu, B.-Q.~Ma,
{Phys. Lett. B} 820 (2021) 136518.

\bibitem{Li:2020uef}
H.~Li, B.-Q.~Ma,
{Sci. Bull.} 65 (2020) 262.

\bibitem{Amelino-Camelia:2016ohi}
G.~Amelino-Camelia, G.~D'Amico, G.~Rosati, N.~Loret,
{Nature Astron.} 1 (2017) 0139.

\bibitem{Ellis:2004ab}
J.~Ellis, N.~E.~Mavromatos, M.~Westmuckett,
{Phys. Rev. D} 70 (2004) 044036;
ibid. 71 (2005) 106006.

\bibitem{Ellis:2008gg}
J.~Ellis, N.~E.~Mavromatos, D.~V.~Nanopoulos,
{Phys. Lett. B} 665 (2008) 412.

\bibitem{Li:2009tt}
T.~Li, N.~E.~Mavromatos, D.~V.~Nanopoulos, D.~Xie,
{Phys. Lett. B} 679 (2009) 407.

\bibitem{Li:2021eza}
C.~Li, B.-Q.~Ma,
{Results Phys.} 26 (2021) 104380.

\bibitem{Ellis:2003ia}
J.~Ellis, N.~E.~Mavromatos, D.~V.~Nanopoulos, A.~S.~Sakharov,
{Int. J. Mod. Phys. A} 19 (2004) 4413;
{Nature} 428 (2004) 386.

\bibitem{Ellis:2003sd}
J.~R.~Ellis, N.~E.~Mavromatos, A.~S.~Sakharov,
{Astropart. Phys.} 20 (2004) 669.

\bibitem{Li:2021duv}
C.~Li, B.-Q.~Ma,
{Sci. Bull.} 66 (2021) 2254.

\end{thebibliography}
\end{document}